\documentclass{ws-p8-50x6-00}

\def\gs{\mathrel{\raise0.35ex\hbox{$\scriptstyle >$}\kern-0.6em
\lower0.40ex\hbox{{$\scriptstyle \sim$}}}}
\def\ls{\mathrel{\raise0.35ex\hbox{$\scriptstyle <$}\kern-0.6em
\lower0.40ex\hbox{{$\scriptstyle \sim$}}}}

\begin{document}

\title{sub-mm counterparts to Lyman-break galaxies
}

\author{Scott Chapman}

\address{Observatories of the Carnegie Institution of
        Washington, Pasadena, CA 91101,~~U.S.A.\\E-mail: schapman@ociw.edu}

\author{Douglas Scott, Colin Borys, Mark Halpern}

\address{Department of Physics \& Astronomy, 
	University of British Columbia,
        Vancouver, B.C.~V6T 1Z1,~~Canada}



\maketitle

\abstracts{ We summarize the main results from our SCUBA survey of
Lyman-break galaxies (LBGs) at $z\sim3$.  Analysis of our sample of LBGs
reveals a mean flux of S$_{850}=$0.6$\pm$0.2\,mJy, while simple models of
emission based on the UV properties predict a mean flux about twice as
large. Known populations of LBGs are expected to contribute flux to the
weak sub-mm source portion of the far-IR background, but are not likely to
comprise the bright source (S$_{850}>5$\,mJy) end of the SCUBA-detected
source count.
The detection of the LBG, Westphal-MM8, at 1.9\,mJy 
suggests that deeper observations of
individual LBGs in our sample could uncover detections at similar levels,
consistent with our UV-based predictions.
By the same token, many sub-mm selected sources
with S$_{850}$$<$2\,mJy could be LBGs.
The data are also consistent with the FarIR/$\beta$ relation holding at $z=3$.
}

\section{Introduction -- Selecting sub-mm luminous LBGs}

The Sub-millimeter Common
User Bolometer Array (SCUBA\cite{scuba}) on the 15-m JCMT\footnote{The
JCMT is operated by the Joint Astronomy Centre on behalf of the United
Kingdom Particle Physics and Astronomy Research Council (PPARC), the
Netherlands Organisation for Scientific Research, and the National
Research Council of Canada.} has sparked a revolution in our 
understanding of dust obscured star formation and active galactic nuclei.
However, going back several years to the initial discovery of sub-mm sources in 
the field\cite{sib}, we were rather naive about
the nature of sub-mm selected sources. It seemed at the time perfectly
reasonable that the high star formation rate tail of the 
known $z\sim3$ 
LBG population should allow us to quickly pre-select and observe
some large fraction of the bright SCUBA sources.
 
The $R$-band for $z\sim3$ galaxies (rest frame UV continuum) provides
a direct measure of the light from young, hot stars. The spectral
slope through the rest frame UV continuum allows a determination
of the extinction of this light due to interstellar dust, and thereby
prescribes a dust correction to the star formation rate (SFR).
Thus the goal of our project was to choose those few LBGs exhibiting a
combination of a bright $R$-magnitude and a steep $g-R$ slope 
(large dust correction) resulting in a sample of galaxies thought to 
have high SFRs ($>200$\,M$_\odot$/yr)
and thus be detectable in the sub-mm with SCUBA. 
It is important to note that these objects were not just a few outliers
with extremely large dust correction factors, but were also amongst the
most intrinsically luminous of the LBG population.

\subsection{An initial sample of high SFR LBGs}
The results of our initial study have been presented in
Chapman et al.~(2000)\cite{chapman2000}, with further implications explored in
Adelberger \& Steidel (2000)\cite{as2000}. From an original sample of 16 LBGs
chosen for followup with SCUBA, only 8 turned out to 
have the sought properties (high expected SFR) upon more detailed optical
analysis. 
Predicting the 850 micron flux density (S$_{850}$)
for LBGs from the UV continuum
was accomplished by first employing the empirical farIR/$\beta$
relation for local starburst galaxies\cite{meurer99}
to estimate the total farIR luminosity from
the UV slope ($\beta$). We then identified the S$_{850}$ point 
through an
empirical measure of the typical spectral energy distribution (SED)
for star forming galaxies in the farIR/sub-mm region, extracted from
a large sample of local LIRGs and ULIRGs\cite{as2000,chapman2000}.
A clear uncertainty in this prescription is the paucity of 
high-$z$ star forming objects with sub-mm measurements
available to validate our prediction recipe.

SCUBA observations during observing runs in winter 1998 yielded
only one clear detection in our sample,
Westphal-MMD11. This object in particular was found
to be rather extraordinary in many of it's properties, and will be
discussed in the following section.
The UV-based predictions of sub-mm flux density were typically too high 
for the sample on the whole, except for Westphal-MMD11 which has
considerably more sub-mm flux than expected.
However, large errors bars and a relatively small sample of objects
made it difficult to extract robust conclusions from the study.
The results of this study are summarized in Figure~1.
However, new evidence as presented below suggests that this
original study may have been a
somewhat misleading result.

\begin{figure}[th]
\centerline{\psfig{file=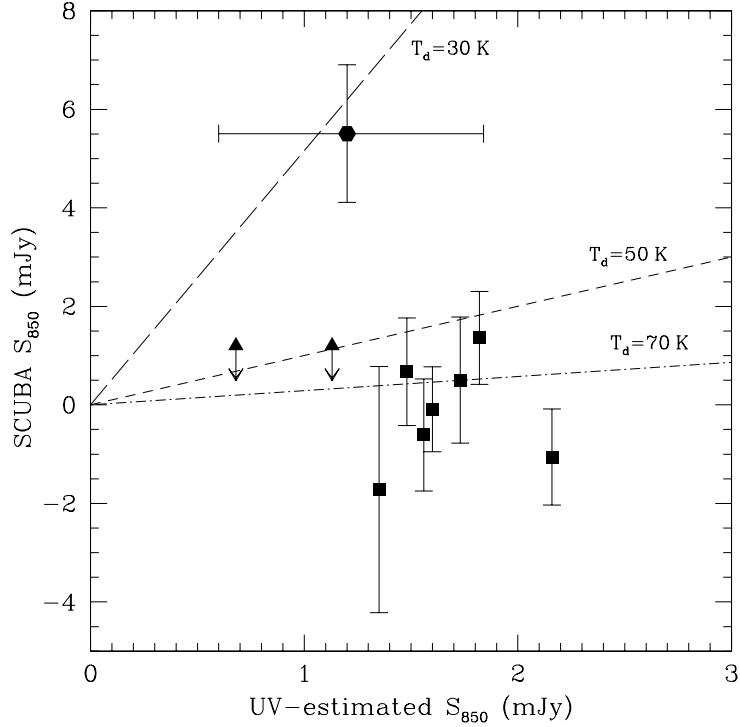,angle=0,width=4.0in}}
\caption{The predicted vs measured sub-mm flux density for our
initial sample of eight LBGs. The dust temperature dependence of our
predictions is shown for 30, 50 and 70\,K. Also plotted are 2 LBG sources
from the HDF (triangles).
}
\end{figure}

\section{Sub-mm detected LBGs at $z\sim3$}

\subsection{Westphal-MMD11}
Westphal-MMD11 currently represents the highest redshift
source ($z=2.98$) detected with SCUBA, which is not an AGN.
Upon subsequent optical study, Westphal-MMD11 
seems more akin to HR10\cite{dey99}, the dusty starbursting
ERO, than what one might consider a local analog of LBGs such as M82.
It is luminous in the near-IR with $R-K=4.5$, almost 2 magnitudes
larger in $R-K$ than the median for the LBG population\cite{shapley2000}.
With S$_{850}=5.5$\,mJy 
and L$_{FIR}$ close to 10$^{13}$h$_{50}^{-2}$\,L$_\odot$,
the question perhaps is why is it visible in the UV at all?
A recent near-IR spectrum taken with NIRSPEC on the Keck 
telescope\cite{ste00}
reveals a double peaked line profile, with continuum
emission only present under one peak. This
suggests geometrical effects involving large amounts of dust are likely 
at work, perhaps similar to the merging ULIRG, Arp220.

\subsection{SMMJ14011}

The lensed galaxy SMM\,J14011+0252\cite{sibk,f99} 
is the only sub-mm selected object clearly identified as a high
redshift ($z=2.565$)
galaxy without an obviously active nuclues\cite{i00}.
Followup optical photometry in a filter set matched to the 
Steidel et al.~(1999)\cite{steidel1999} LBG survey\cite{as2000} 
revealed that the restframe-UV SED of SMMJ14011
is actually indistinguishable from Westphal-MMD11, yet their optical SEDs
are quite different.
Within the SMM\,J14011+0252
system, interferometry observations\cite{f99} reveal that the dominant
sub-mm emission resides in a red component (J1), with the dominant UV
emission appearing in a neighboring component (J2).
At face value, this questions the validity of the
unobscured star formation rate prediction based on the UV.
The UV properties of a spatially separated component should have little
bearing on the dust and star formation in J1.
On the other hand the empirical correlation between $\beta$ and  
$L_{bol,dust}/L_{UV}$ is rather mysterious -- it is difficult to
understand why this
correlation should exist at all. But local observations suggest that it
does hold,
and the scatter in the relation may relate a variety of processes involved
in starbursting galaxies at various evolutionary stages.

\subsection{Westphal-MM8}
Identified  as a LBG in the surveys of
Steidel et al.~(1999)\cite{steidel1999}, Westphal-MM8
possesses  a rest-frame  UV  spectrum  which indicates
an exceptionally large SFR ($>100 M_\odot /yr$ ).
However, the infrared colors
and overall properties do not single this object out as unusual
for the population in any other respects.
Although still a relatively extreme object, Westphal-MM8 
represents one of the faintest
galaxies ever detected with SCUBA (S$_{850}=1.9\pm0.5$\,mJy), 
comparable to the faintest objects in
the Hubble Deep Field SCUBA observation of Hughes et al.~(1998).
Simply scaling the  observed submillimetre photometry from that of Arp220
implies that MM8 has  an infrared  luminosity  of ${\rm
L_{IR}\sim4\times10^{12}\  h_{50}^{-2} L_\odot}$.
Although original results\cite{chapman2000,as2000}
were somewhat pessimistic about the sub-mm detection of LBGs,
this measurement of MM8 gives us renewed confidence that
we may be able to pre-select those extreme LBGs emitting in the
S$_{850}$=1-2\,mJy range.

\section{A larger sample of high-SFR LBGs}
The case of Westphal-MM8 reveals the inherent difficulty in detecting
even the most sub-mm luminous members of the LBG population with SCUBA.
Clearly spending two JCMT shifts per object is not an option for 
studying the properties of a large sample of LBGs. 

A new and larger sample consisting of 33 LBGs has now been observed
to an average 
RMS $\sim$1.3\,mJy with SCUBA. The sample breaks down as follows:
8 high SFR LBGs from our original sample,
12 red LBGs ($R-K>3.5$) which could be `like' W-MMD11, and
13 new LBGs similar to the original sample of large predicted SFR.
Some of these objects are selected from fields with better photometry
than the original surveys, and photometric errors could conceivably
be less of an uncertainty.
Still, only 2 clear detections emerge (W-MMD11 and W-MM8), although
certain sub-samples exhibit clear detections in their average flux
density.

We summarize the average properties of this new sample as follows:

S$_{850}$ = 0.6$\pm$0.2\,mJy 

S$_{450}$ = 2.9$\pm$2.2\,mJy

Our results clearly indicate that LBGs do not form a significant part
of the bright sub-mm selected sources (S$_{850}>5$\,mJy). 
These numbers include MMD11, as the numbers are small enough that
we do not yet know if MMD11 is highly unusual, or merely the tail end
of the red LBG population. Several other red LBGs are marginally
detected with S$_{850}\sim3$\,mJy.
The S$_{850}$ predicted from the UV for this sample
is approximately 1.2\,mJy and is
thus not drastically in conflict with the measured flux, but may
indicate a downwards correction of a factor $\sim$2$\times$ in 
L$_{\rm dust,bol}$ for the most luminous LBGs.
This result indicates that the FarIR/$\beta$ relation appears at
least consistent at $z=3$ with what we observe locally.

\subsection{Far-IR background}

For an $\Omega=0.3$, $\Lambda=0.7$ universe, the average dust luminosity 
of our SCUBA observed LBG sample is 
L$_{bol}$(dust) = 2.4$\pm$0.8$\times10^{11}$\,L$_\odot$.
The average $R$(AB)=23.9 implies L$_{UV}$ = 4.6$\times10^{10}$\,L$_\odot$. 
The average {\it obscuration} for the sample, $<L_{dust} / L_{UV}>$ 
= $4.8\pm1.6$.
UV-based predictions\cite{as2000} requre $<L_{dust} / L_{UV}>=6$   
to recover the bulk of the 850 micron background, dominated by 0.5--2\,mJy
sources.

Although submm sources $>5$mJy appear largely distinct from the LBG 
population\cite{bcr,crl}, 
they form a
relatively small amount of FarIR Background. However, deep sub-mm 
observations in the HDF\cite{hughes1998,peacock2000}
suggest that $S_{850}\sim2$mJy sources are observed which have no 
obvious optical counterparts and would not form part of typical ground
based optical surveys. 
Thus it remains to be seen just what properties the bulk of 
these crucial $S_{850}\sim1-2$mJy sources actually have.



%
%
%
%
\section*{Acknowledgments}
We thank our collaborators, 
C. Steidel, K. Adelberger (Caltech)
S. Morris (DAO), M. Pettini (Cambridge)
M. Dickinson, M. Giavalisco (STSCI) for helping to put all the pieces 
together in this complicated puzzle.
SCC acknowledges travel support from Carnegie Observatories 
to attend this conference.


\begin{thebibliography}{99}
\bibitem{scuba} Holland, W.S., et al., \Journal{MNRAS}{303}{659}{1999}.
\bibitem{sib} Smail, I., Ivison, R.J., Blain, A.W., \Journal{ApJ}{490}{L5}{1997}.
\bibitem{chapman2000} Chapman S.C., Scott D., Steidel C. et al., \Journal{MNRAS}{in press}{2000}.
\bibitem{as2000} Adelberger, K.L., Steidel, C.C., \Journal{ApJ}{$\!\!$}{in press}{2000}.
\bibitem{meurer99} Meurer G.\,R., 
	Heckman T.\,M., Calzetti D., \Journal{ApJ}{521}{64}{1999}
\bibitem{dey99} Dey, A., et al., \Journal{ApJ}{519}{610}{1999}
\bibitem{shapley2000} Shapley A., et al., \Journal{ApJ}{$\!\!$}{in preparation}{2000}
\bibitem{ste00} Steidel, C.C., et al., \Journal{ApJ}{$\!\!$}{in preparation}{2000}
\bibitem{sibk} Smail, I., Ivison, R.J.,  Blain, A.W., Kneib, J.-P., \Journal{ApJ}{507}{L21}{1998}.
\bibitem{f99} Frayer, D.T., et al., \Journal{ApJ}{514}{L13}{1999}.
\bibitem{i00} Ivison, R.J., et al., \Journal{MNRAS}{315}{209}{2000}.
\bibitem{steidel1999} Steidel, C.C., et al. \Journal{ApJ}{519}{1}{2000}.
\bibitem{bcr} Barger A.J., Cowie L, Richards E., \Journal{AJ}{119}{2092}{2000}.
\bibitem{crl} Chapman S.C., Richards E., Lewis G., 
	\Journal{Nature}{$\!\!$}{submitted}{2000}.
\bibitem{hughes1998} Hughes, D.H., et al., \Journal{Nature}{394}{241}{1998}.
\bibitem{peacock2000} Peacock J., et al., \Journal{MNRAS}{$\!\!$}{in press}{2000}.

\end{thebibliography}
\end{document}